\documentclass[12pt]{article}
\usepackage{epsfig}
\begin{document}
\setlength{\baselineskip}{0.30in}
\newcommand{\nc}{\newcommand}
\newcommand{\beq}{\begin{equation}}
\newcommand{\eeq}{\end{equation}}
\newcommand{\be}{\begin{eqnarray}}
\newcommand{\ee}{\end{eqnarray}}
\newcommand{\num}{\nu_\mu}
\newcommand{\nue}{\nu_e}
\newcommand{\nut}{\nu_\tau}
\newcommand{\nus}{\nu_s}
\newcommand{\mnus}{M_s}
\newcommand{\taus}{\tau_{\nu_s}}
\newcommand{\nnt}{n_{\nu_\tau}}
\newcommand{\rnt}{\rho_{\nu_\tau}}
\newcommand{\mnt}{m_{\nu_\tau}}
\newcommand{\tnt}{\tau_{\nu_\tau}}
\newcommand{\bi}{\bibitem}
\newcommand{\rar}{\rightarrow}
\newcommand{\lar}{\leftarrow}
\newcommand{\lrar}{\leftrightarrow}
\newcommand{\dm}{\delta m^2}
\newcommand{\so}{\, \mbox{sin}\Omega}
\newcommand{\co}{\, \mbox{cos}\Omega}
\newcommand{\sotil}{\, \mbox{sin}\tilde\Omega}
\newcommand{\cotil}{\, \mbox{cos}\tilde\Omega}
\makeatletter
\def\alt{\mathrel{\mathpalette\vereq<}}
\def\vereq#1#2{\lower3pt\vbox{\baselineskip1.5pt \lineskip1.5pt
\ialign{$\m@th#1\hfill##\hfil$\crcr#2\crcr\sim\crcr}}}
\def\agt{\mathrel{\mathpalette\vereq>}}

\newcommand{\eq}{{\rm eq}}
\newcommand{\tot}{{\rm tot}}
\newcommand{\M}{{\rm M}}
\newcommand{\coll}{{\rm coll}}
\newcommand{\ann}{{\rm ann}}
\makeatother

\begin{center}
\vglue .06in
{\Large \bf {Heavy sterile neutrinos:
Bounds from big-bang nucleosynthesis and
SN~1987A}}
\bigskip
\\{\bf A.D. Dolgov
\footnote{Also: ITEP, Bol. Cheremushkinskaya 25, Moscow 113259, Russia.}
\footnote{e-mail: {\tt dolgov@fe.infn.it}},
S.H. Hansen\footnote{e-mail: {\tt sthansen@fe.infn.it}}
 \\[.05in]
{\it{INFN section of Ferrara\\
Via del Paradiso 12,
44100 Ferrara, Italy}
}}
\\{\bf G. Raffelt \footnote{e-mail: {\tt raffelt@mppmu.mpg.de}}} \\
{\it{Max-Planck-Institut f\"ur Physik (Werner-Heisenberg-Institut)\\
F\"ohringer Ring 6, 80805 M\"unchen, Germany
}}
\\{\bf D.V. Semikoz \footnote{e-mail: {\tt semikoz@ms2.inr.ac.ru}}} \\
{\it{Max-Planck-Institut f\"ur Physik (Werner-Heisenberg-Institut)\\
F\"ohringer Ring 6, 80805 M\"unchen, Germany\\
and\\
Institute of Nuclear Research of the Russian Academy of Sciences\\
60th October Anniversary Prospect 7a, Moscow 117312, Russia}}
\\[.40in]
\end{center}

\begin{abstract}
Cosmological and astrophysical effects of heavy (10--200~MeV) sterile 
Dirac neutrinos, mixed with the active ones, are considered. The bounds on 
mass and mixing angle from both supernovae and big-bang nucleosynthesis
are presented.
\end{abstract}    

\begin{center}
\end{center}

\newpage

\section{Introduction}

There are strong and well known indications that the three ordinary
(active) neutrinos $\nu_a$ ($a=e$, $\mu$, $\tau$) are mixtures of
different mass eigenstates $\nu_j$, with $j=1,2,3,\ldots$ (for a
review see e.g.~\cite{rev}).  Active neutrinos can mix among
themselves, and possibly {\it only\/} among themselves, in which case
$j=1,2,3$. However, it is also possible that there are mixings of the
active neutrinos with new sterile ones, $\nu_s$, that do not have any
direct coupling to $W$ and $Z$ bosons mediating weak interactions.
Sterile neutrinos have repeatedly been suggested as solutions of
various anomalies in neutrino experiments.  In particular, if all the
present day data are correct, then at least one sterile neutrino must
participate.  For an explanation of the known neutrino anomalies, the
mass differences should be very small, it should lie in the eV range
or even in the sub-eV range.  However, if we admit that there may be
new neutrino states mixed to the usual ones, we should keep an open
mind to possible values of their masses. It is an interesting question
which values of masses and mixing angles can be excluded by direct
experiments as well as by cosmology and astrophysics. The last two
topics are addressed in the present paper which was partly stimulated
by the recent limits on mixing of $\nut$ with a heavy sterile neutrino
found by the NOMAD collaboration~\cite{nomad}, and partly by our
paper~\cite{dolgov00} where we have found that cosmology and
astrophysics practically exclude the interpretation of the KARMEN
anomaly~\cite{karmen} by a 33.9~MeV neutrino mixed with $\nut$ 
(astrophysical and cosmological limits on 33.9 MeV sterile neutrino
were also considered in ref.~\cite{barger95}).
According to a statement of the KARMEN collaboration made at Neutrino
2000~\cite{karmen2000} the anomaly 
was not observed in upgraded detector
KARMEN 2, but the question
still remains which area in the mass-mixing-plane can be excluded.

In what follows we will derive the bounds on masses and mixings that
follow from big bang nucleosynthesis (BBN) and from the duration of
the supernova (SN)~1987A neutrino burst. We essentially repeat our
previous work~\cite{dolgov00}, but now lifting the restriction on a
specific value of mass and lifetime of the heavy neutrinos and
considering mixings with all active states $\nu_e$, $\nu_\mu$ and
$\nu_\tau$. We assume that the heavy neutrinos have Dirac mass. In
the case of Majorana mass the limits would be slightly weaker.

\section{Preliminaries}

We assume that the sterile neutrino mixes predominantly with only one
active flavour $\nu_a=\nue$, $\num$ or $\nut$.  The mixed flavour states
are expressed in terms of the mass eigenstates and the
mixing angle $\theta$ as
\be
\nu_a &=& \cos\theta ~\nu_1 + \sin \theta ~\nu_2\,,  \nonumber \\
\nu_s &=& -\sin \theta ~\nu_1 + \cos\theta ~\nu_2\,,
\label{nuas}
\ee
where $\nu_1$ and $\nu_2$ are assumed to be the light and heavy mass
eigenstates, respectively.

This mixing couples the heavier neutrino to the $Z^0$, allowing for
the decay channel
\be
\nu_2 \rightarrow \nu_1 + \ell + \bar \ell \, ,
\label{dec}
\ee
where $\nu_1$ is mostly an active flavour and $\ell$ is any lepton with
a mass smaller than the mass $m_2$ of the heavy neutrino.  We assume
that $m_2 < 2 m_\mu$ so that the decay into $\bar\mu \mu$ and
$\bar\tau \tau$ is kinematically forbidden. If the active neutrino
mixed with $\nus$ is either $\num$ or $\nut$, then we can translate
the mixing angle into lifetime as
\be
\tau_{\nu_s}  \equiv \Gamma_{\nu_2}^{-1} = { 
1.0~{\rm sec} \over (M_s/\mbox{10 MeV})^5\, \mbox{sin}^2 2\theta}  \, .
\label{taumu}
\ee
For the mixing with $\nue$ the numerator is 0.7~sec; it is is
different because of the presence of charged-current interactions.

A sterile neutrino mixed with $\nu_a$ could be observed in direct
experiments, in particular where upper bounds on neutrino masses are
obtained (see the list of references in~\cite{pdg}). The most accurate
limits of course exist for $\nue$~\cite{lobashev99,weinheimer99},
roughly $m_{\nue} <3$~eV.  However, these experiments cannot help much
in eliminating a heavy sterile neutrino because they are not sensitive
to the mass range $M_s > 10$~MeV which we consider. Such heavy
neutrinos are simply not produced in beta-decays and their impact is
only indirect, e.g.~they renormalize vector and axial coupling
constants.

The upper limit on the tau-neutrino mass obtained by the ALEPH
Collaboration~\cite{aleph} could be translated into limits on
mass/mixing of $\nut$ with $\nus$.  However, one would need to
reanalyse the data under the assumption of two (or several) mixed
neutrinos, taking account of the sensitivity to measure the energy
spectrum for different values of $m_\nu$.  The bound obtained in
Ref.~\cite{ahluwalia97} based on the assumption that the average
neutrino mass $\langle m \rangle = \cos^2\theta~ m_1 + \sin^2\theta~
m_2$ should be smaller than the experimental upper limit is too naive
and can serve only as a rough order-of-magnitude estimate. Moreover
the NOMAD bounds~\cite{nomad} are much more restrictive and thus we
will not pursue the subject of the ALEPH bounds here.

\section{Cosmological production and freeze-out of 
heavy sterile neutrinos}

In the early universe sterile neutrinos are produced through their
mixing with the active ones. The production rate for relativistic
$\nus$ (i.e.~for $T_\gamma \geq m_2$) can be approximately estimated as~\cite{barbieri90}:
\be
{\Gamma_s \over H} \approx {\sin^2 2 \theta_\M \over 2} 
\left( {T_\gamma \over T_w }\right)^3 \, ,
\label{gammas}
\ee
where $H$ is the Hubble expansion parameter, $T_\gamma$ is the plasma temperature 
equal to the photon temperature.  
$T_w$ is the decoupling
temperature of active neutrinos, taken to be 3~MeV, and $\theta_\M$ is
the the mixing angle in the medium.  According to the calculations of
Ref.~\cite{notzold88} one finds in the limit of small mixing:
\be
\sin 2 \theta_\M \approx {\sin 2\theta \over
1+ 0.76 \times 10^{-19}\, (T_\gamma/\mbox{MeV})^6 (\delta m^2/\mbox{MeV}^2)^{-1}}~,
\label{thetam}
\ee
The matter effects become essential for
\be 
T_\gamma>1.5\times 10^3\,\mbox{MeV} (\dm/\mbox{MeV}^2)^{1/6} .
\label{tmatt}
\ee

For $\Gamma_s/H > 1$, sterile neutrinos were abundantly produced and
their number density was equal to that of light active neutrinos, at
least during some epoch.  The production rate reaches a maximum at
$T_{\rm max} = 1.28\times10^3 (\dm/\mbox{MeV}^2)^{1/6}$~MeV. For the
masses that are considered below, $T_{\rm max}$ is well above the
neutrino mass.

If the equilibrium number density of sterile neutrinos is reached, it
would be maintained until $T_f \approx 4 (\sin 2\theta)^{-2/3}$~MeV.
This result does not depend on the heavy neutrino mass because they
annihilate with massless active ones, $\nu_2 + \nu_a
\rar all$. The heavy neutrinos would be relativistic at decoupling and
their number density would not be Boltzmann suppressed if, say, 
$T_f>M_s/2$. This gives  
\be 
\sin^2 2\theta (\dm/\mbox{MeV}^2)^{3/2} < 500~.
\label{relat}
\ee
 If this condition 
is not fulfilled the impact of $\nu_s$ on BBN
would be strongly diminished. On the other hand, for a sufficiently
large mass and non-negligible mixing, the $\nu_2$ lifetime given in
Eq.~(\ref{taumu}) would be quite short, so that they would all decay
prior to the BBN epoch.  (To be more exact, their number density would
not be frozen, but follow the equilibrium form $\propto e^{-M_s/T_\gamma}$.)

Another possible effect that could diminish the impact of heavy
neutrinos on BBN is entropy dilution. If $\nu_2$ were decoupled while
being relativistic, their number density would not be suppressed
relative to light active neutrinos. However, if the decoupling
temperature is higher than, say, 50~MeV pions and muons were still
abundant in the cosmic plasma and their subsequent annihilation would
diminish the relative number density of heavy neutrinos. If the
decoupling temperature is below the QCD phase transition the dilution
factor is at most $17.25/10.75 =1.6$.  Above the QCD phase transition
the number of degrees of freedom in the cosmic plasma is much larger
and the dilution factor is approximately 5.5. However, these effects
are essential for very weak mixing, for example the decoupling
temperature would exceed 200~MeV if $\sin^2 2\theta < 8\times
10^{-6}$. For such a small mixing the life-time of the heavy
$\nu_2$ would exceed the nucleosynthesis time and they would be
dangerous for BBN even if their number density is 5 times diluted.

 Sterile neutrinos would never be abundant in the
universe if $\Gamma_s/H < 1$. In fact we can impose a stronger
condition demanding that the energy density of heavy neutrinos
should be smaller than the energy density of one light neutrino
species at BBN ($T\sim 1$ MeV). Taking into account a possible entropy
dilution by factor 5  we obtain the 
bound:
\be
\left( \dm/\mbox{MeV}^2\right)\,\sin^2 2\theta 
< 2.3\times 10^{-7} \, .
\label{dmsin}
\ee
Parameters satisfying this conditions cannot be excluded by BBN.  A
more detailed consideration permits one to impose a somewhat better
limit, but we will not go into such detail here.

Keeping these restrictions in mind we proceed to derive bounds on
$M_s$ and the mixing angle $\theta$ demanding that the influence of
heavy neutrinos on BBN should not be too strong.

\section{Big-Bang Nucleosynthesis}

In general, if a neutrino has a mass exceeding an MeV then it may
influence the light element abundances unless it decays much before
the onset of nucleosynthesis (see e.g.~\cite{dolgov98,dolgov99} and
references therein).  A heavy unstable sterile neutrino will affect
BBN in several ways. The main effect is through the increased energy
density which leads to a faster expansion and hence an earlier
freeze-out of the $n/p$-ratio.  However, also the decay products must
be taken into account since the electron neutrinos directly enter the
$n$-$p$ reactions. Moreover, if the $\nus$ decays into the $e^+e^-$
channel the temperature evolution is altered
(see~\cite{dolgov00,dolgov97} for discussion).

The calculations are described in detail in Ref.~\cite{dolgov00}; here
we only briefly describe the basic steps.  First we introduce the
dimensionless variables
\be
x =  m_0 a(t)\quad {\rm and}\quad y = p\, a(t)~,
\ee
where $a(t)$ is the cosmic scale factor and $m_0$ is an arbitrary
normalisation factor that we have chosen as $m_0 = 1$~MeV. We also
introduce a dimensionless temperature $T=T_\gamma/m_0$ and measure
all masses in units of $m_0$.  In terms of these variables the
Boltzmann equations describing the evolution of the sterile neutrino,
$\nus$, and the active neutrinos, $\nu_a$, can be written as
\be
\partial _x f_{\nus} (x,y) &=& \frac{ \left( f_{\nus} ^{eq} - 
f_{\nus} \right) }{\taus/{\rm  sec}} {1.48\, x \over (Tx)^2} 
\left( {10.75 \over g_* (T) }\right)^{1/2}
\nonumber\\
&&\left[ {M_s\over E_{\nus}} 
+ {3\times 2^7 T^3 \over \mnus^3 } \left( \frac{3 
\zeta(3)}{4}
+ \frac{7 \pi^4}{144} \left( {E_{\nus} T\over \mnus^2} + 
{p_{\nus}^2 T \over 3 
E_{\nus}\mnus^2} \right) \right) \right]~,
\label{eqnus} \\
\noalign{\bigskip}
\partial_x \delta f_{\nu_a} &=& \mbox{D}_a (x,y,\mnus) +
\mbox{S}_a(x,y)\, ,
\label{eqnua}
\ee
where $f^{eq}_{\nus} = (e^{E/T} +1)^{-1}$, $\delta f_{\nu_a} =
f_{\nu_a} - (e^y+1)^{-1}$, and $E_{\nus}=\sqrt{\mnus^2 +(y/x)^2}$ and
$p_{\nus}= y/x$ are the energy and momentum of $\nus$. Further,
$g_*(T) = \rho_{\tot} /(\pi^2 T_\gamma^4 /30)$ is the effective number
of massless species in the plasma determined as the ratio of the total
energy density to the equilibrium energy density of one bosonic
species with temperature $T_\gamma$.

The scattering term in Eq.~(\ref{eqnua}) comes from interactions of
active neutrinos between themselves and from their interaction with
electrons and positrons.  For $\nu_e$ it has the form
\be
\mbox{S}_{\nue}(x,y)&&=
0.26 \left({10.75 \over g_*}
\right)^{1/2} (1+g_L^2+g_R^2) (y/ x^4)\nonumber\\
&&{}\times\biggl\{- \delta f_{\nue} 
+{2\over 15}\,{e^{-y}
\over 1+g_L^2+g_R^2} \, [1 +0.75(g_L^2+g_R^2)]\nonumber\\
&&\hskip2em{}\times
\left[ \int dy_2 y^3_2 \delta f_{\nue} (x,y_2)
+{1 \over 8}\, 
\int dy_2 y^3_2 \left( \delta f_{\num} (x,y_2)+\delta f_{\nut}(x,y_2)
\right) \right] \nonumber \\
&&\hskip2em{}+ {3\over 5}(T \cdot x - 1){g_L^2+g_R^2 \over 1+g_L^2+g_R^2} e^{-y}
\left( 11y/12 -1\right) \biggr\}~,
\label{scatt}
\ee
where $g_L = \sin^2 \theta_W +\frac{1}{2}$, while $g_R =\sin^2
\theta_W$.  The corresponding term for $\nu_\mu$ and $\nu_\tau$ is given
by Eq.~(\ref{scatt}) with the exchange $g_L \rightarrow {\tilde g}_L =
\sin^2 \theta_W - \frac{1}{2}$.

The decay term in Eq.~(\ref{eqnua}) comes from the decay of heavy
sterile neutrino according to reactions Eq.~(\ref{dec}). This term
depends on the mixing channel. For $\nu_\tau \leftrightarrow
\nu_s$ mixings we have
\be
\mbox{D}_{\nue,\num}(x,y) &=& { 467\over \mnus^3\taus x^2 }\left({10.75 \over g_*}
\right)^{1/2}
\left( 1-{16y\over 9 M_s x}\right) \left( n_{\nus} - n_{\nus}^{\eq} \right)
\theta \left(M_s x/2-y\right) ~,
\label{snuem} \\
\mbox{D}_{\nut}(x,y)&=&{ 935\over \mnus^3\taus x^2 }\left({10.75 \over g_*} \right)^{1/2}
\left[ 1-{16y\over 9 M_s x} +{2\over 3}\left(1+{\tilde g}_L^2 +g_R^2\right)
\left(1 - {4y \over 3 M_s x} \right)
\right] \nonumber \\
&&{}\times 
\left( n_{\nus} -n_{\nus}^{\eq}\right) \theta \left(M_s x/2-y\right)~,
\label{snut}
\ee
where $n_{\nus}(x)$ is the number density of $\nus$ and $\theta (y)$
is the step function which ensures energy conservation in the decay.
For $\nu_\mu \leftrightarrow \nu_s$ mixing these terms come from
Eq.~(\ref{snuem}) and Eq.~(\ref{snut}) by exchange of 
$\nu_\mu \leftrightarrow \nu_\tau$.
 
For $\nu_e \leftrightarrow \nu_s$ mixing those terms are
\be
\mbox{D}_{\nut,\num}(x,y) &=& {341 \over \mnus^3 \taus x^2 }\left({10.75 \over g_*}
\right)^{1/2}
\left( 1-{16y\over 9 M_s x}\right) \left( n_{\nus} - n_{\nus}^{\eq} \right)
\theta \left(M_s/2-y/x\right),
\label{snuem_e} \\
\mbox{D}_{\nue}(x,y)&=&{683 \over \mnus^3\taus  x^2 }\left({10.75 \over g_*} \right)^{1/2}
\left[ 1-{16y\over 9 M_s x} +{2\over 3}\left(1+g_L^2 +g_R^2\right)
\left(1 - {4y \over 3 M_s x} \right)
\right] \nonumber \\
&&{}\times 
\left( n_{\nus} -n_{\nus}^{\eq}\right) \theta \left(M_s /2-y/x\right)~,
\label{snut_e}
\ee

More details on the derivation of Eqs.~(\ref{eqnus},\ref{eqnua}) 
can be found in  Ref.~\cite{dolgov00}.

The temperature evolution is governed by the equation for energy
conservation:
\be
x\, \frac{d \rho}{dx} = -3 (\rho + p) \, ,
\label{energ}
\ee
where $\rho$ and $p$ are energy density and pressure.

\begin{figure}
\begin{center}
\psfig{file=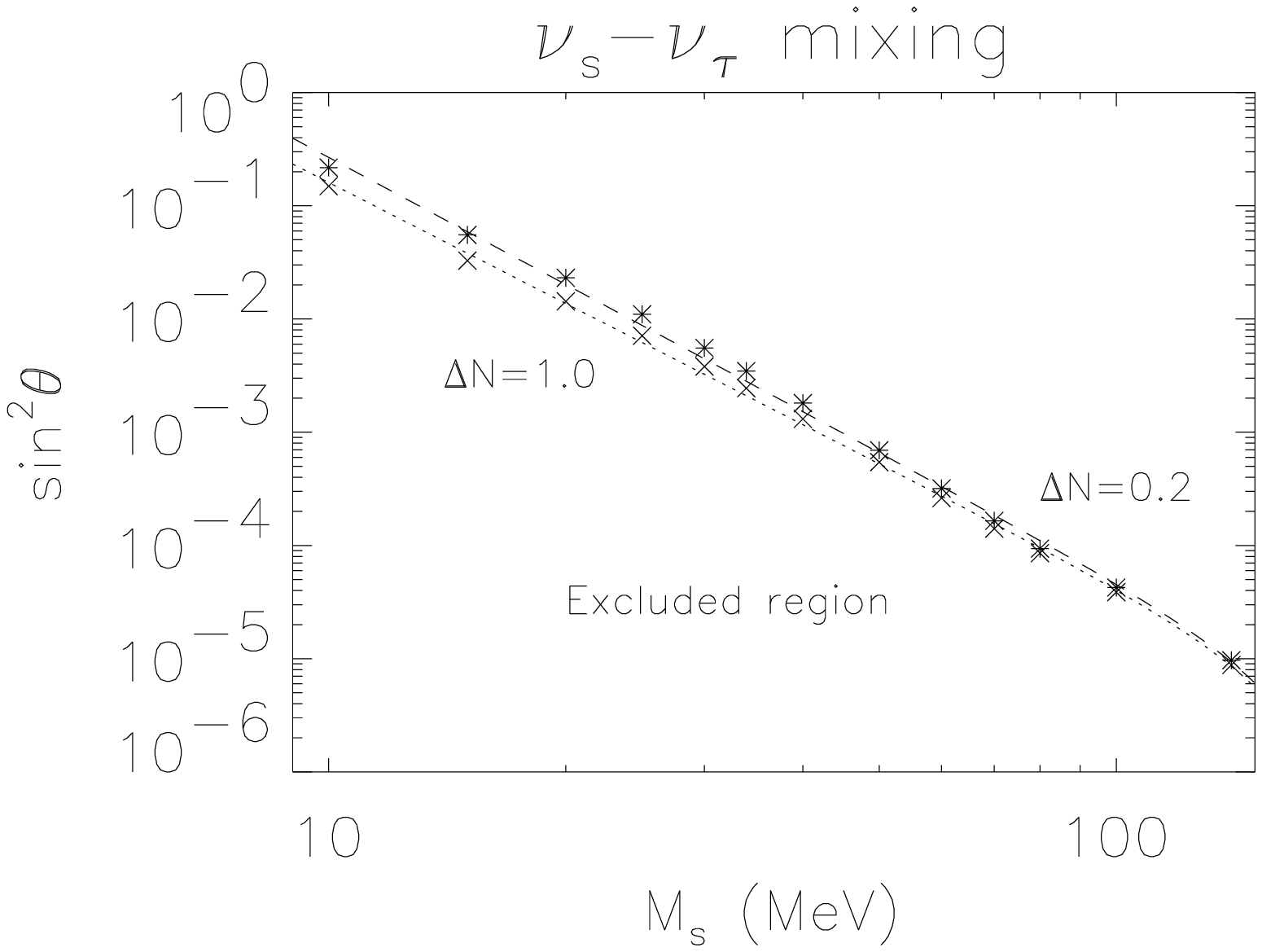,width=4.5in,height=3in}
\psfig{file=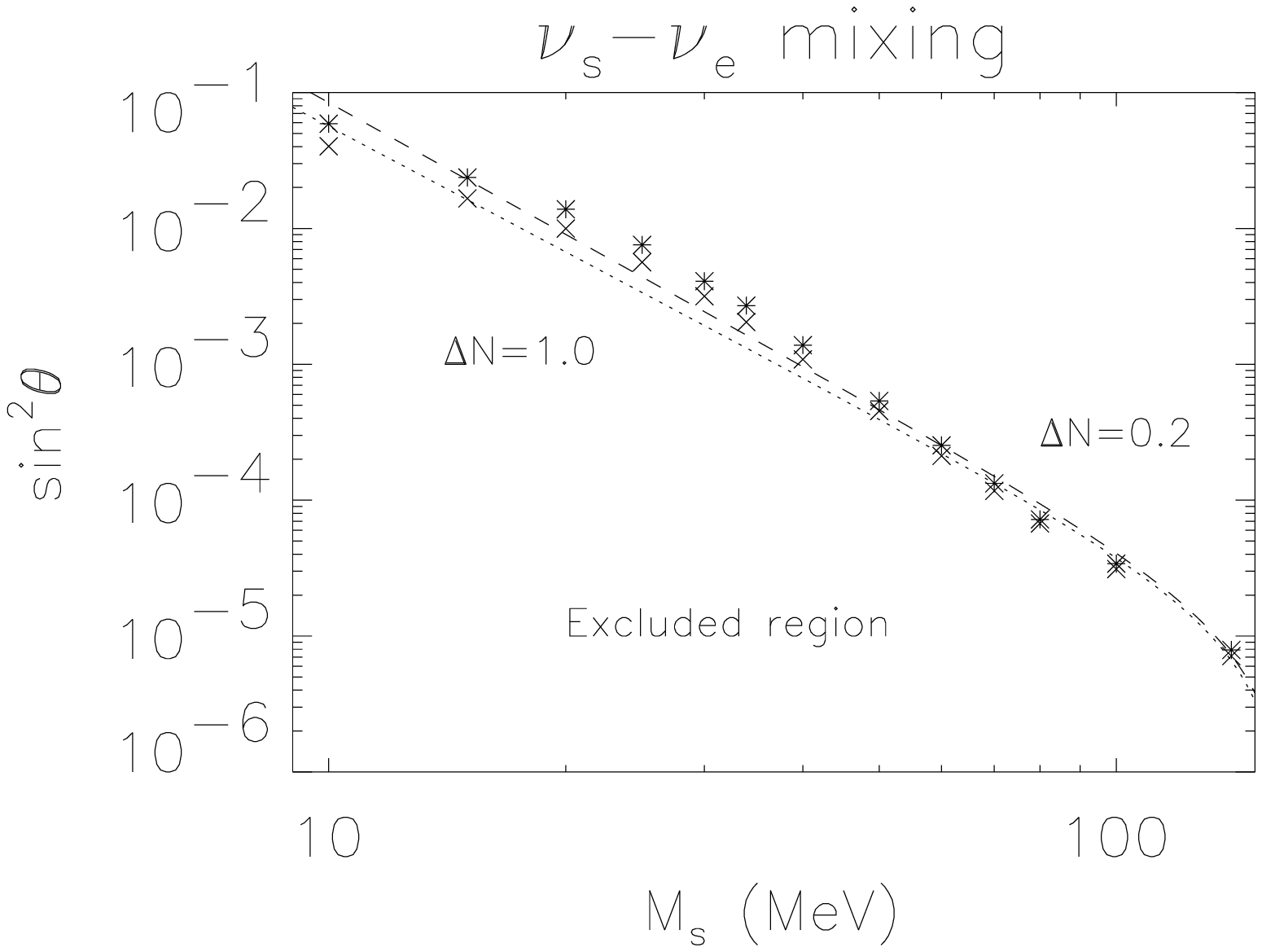,width=4.5in,height=3in}
\caption{Minimum mixing angle between sterile and
active neutrinos, allowed by BBN, as a function of the heavy neutrino
mass, both for an optimistic bound, $\Delta N =0.2$, and for a
conservative bound, $\Delta N = 1.0$. The upper panel corresponds to
$\nu_\mu \leftrightarrow \nu_s$ or $\nu_\tau \leftrightarrow \nu_s$
mixings, while the lower one to $\nu_e \leftrightarrow \nu_s$ mixing.}
\label{fig:smax1}
\end{center}
\end{figure}

\begin{figure}
\begin{center}
\psfig{file=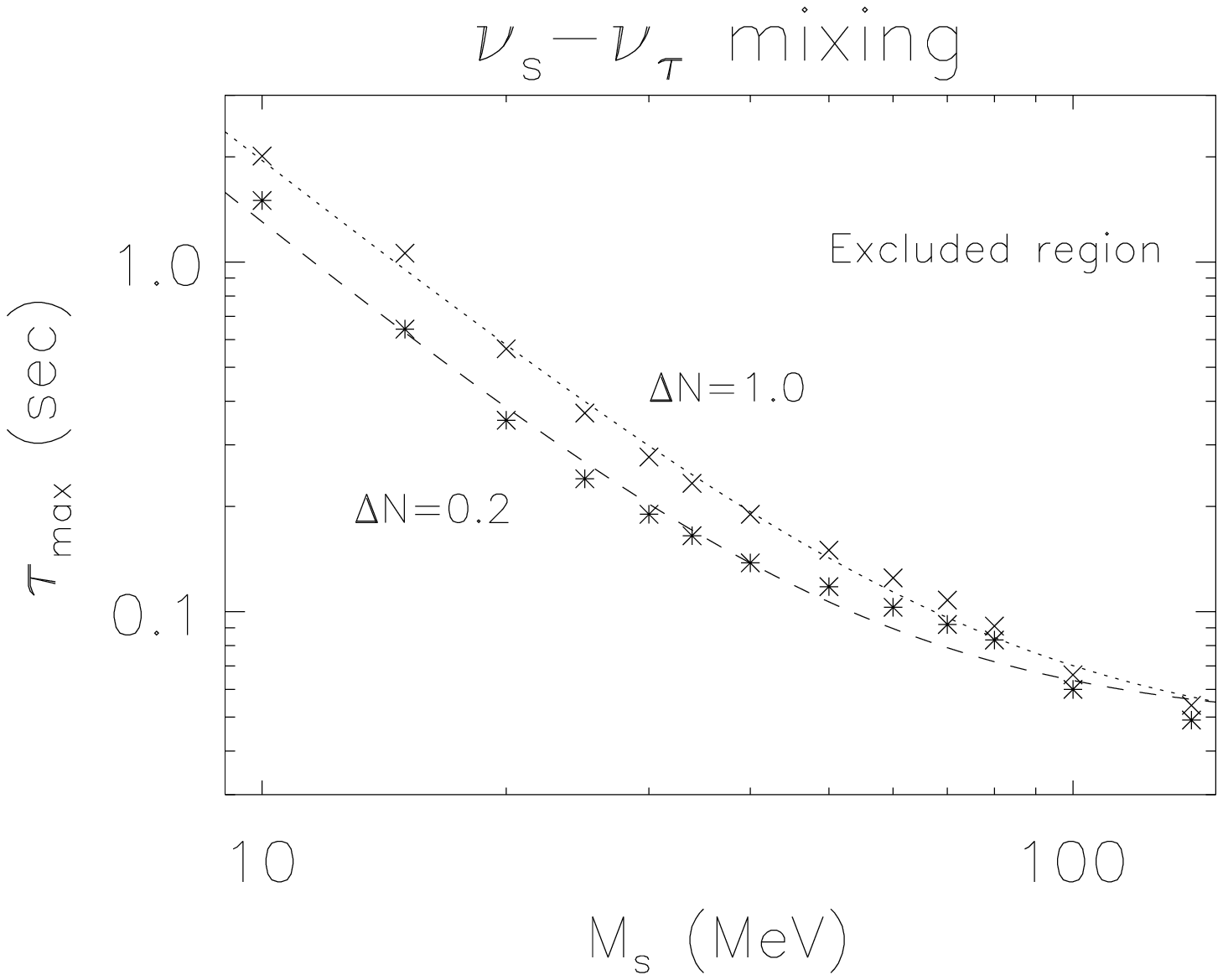,width=4.5in,height=3in}
\psfig{file=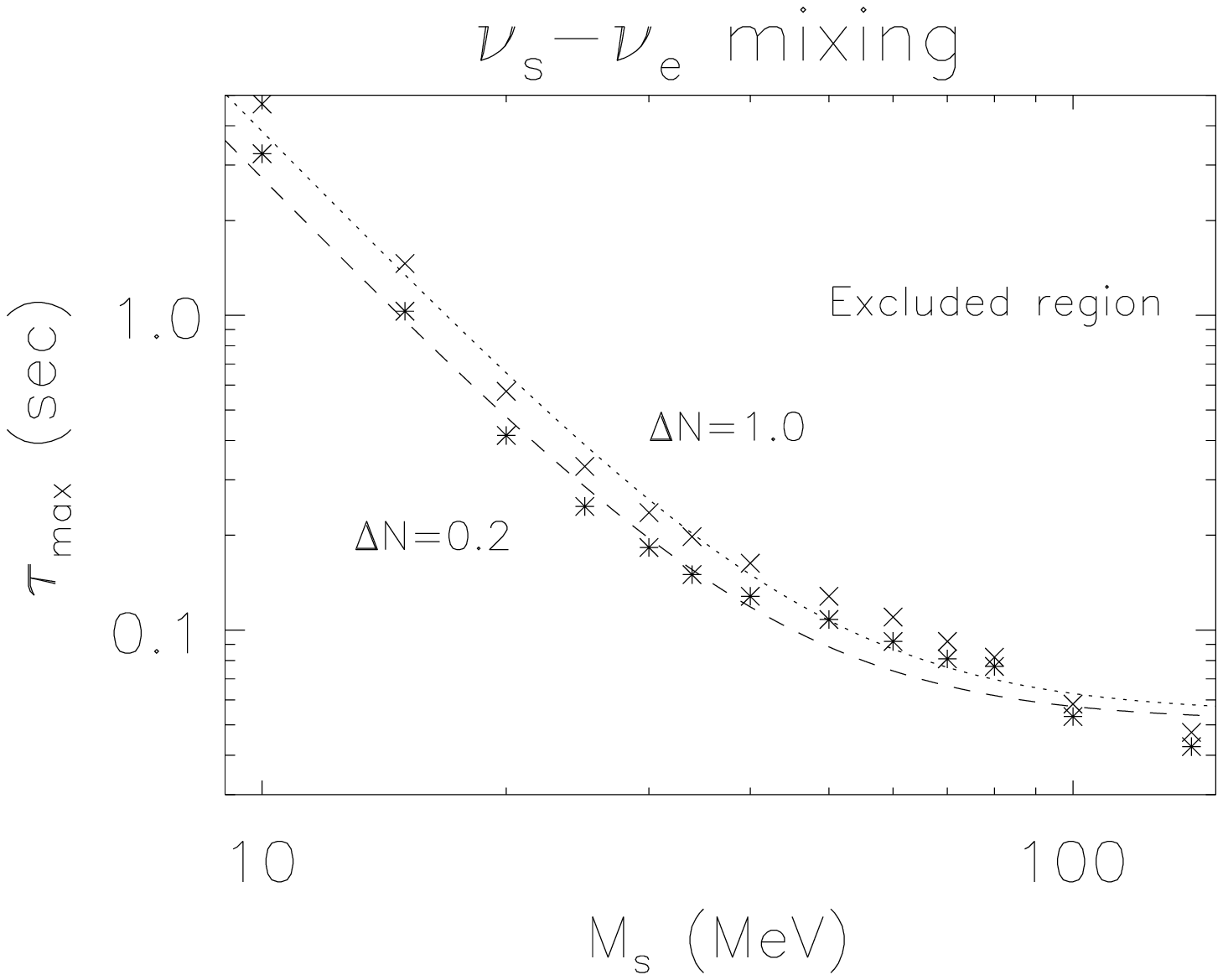,width=4.5in,height=3in}
\caption{
Maximum lifetime of sterile neutrino, allowed by BBN,
as a function of the heavy neutrino
mass, both for an optimistic bound, $\Delta N =0.2$, and for a
conservative bound, $\Delta N = 1.0$. The upper panel corresponds to
$\nu_\mu \leftrightarrow \nu_s$ or $\nu_\tau \leftrightarrow \nu_s$
mixings, while the lower one to $\nu_e \leftrightarrow \nu_s$ mixing.}
\label{fig:tmax1}
\end{center}
\end{figure}

For the solution of Eqs.~(\ref{eqnus},\ref{eqnua},\ref{energ}) we
divide the time region into 3 parts. At early times, when the light
neutrinos are kept in tight equilibrium with the electromagnetic
plasma, we need only to solve the equation for the heavy sterile
neutrinos, Eq.~(\ref{eqnus}). At intermediate temperatures we solve
the full set of equations, taking into account the changed temperature
evolution and the non-equilibrium distribution functions of the
neutrinos.  At late times, long after the neutrinos have decoupled and
the heavy neutrinos have disappeared, we solve only the kinetic
equations governing the $n$-$p$ reactions needed for the
nucleosynthesis code. We use a 800 point grid in the momentum region
big enough to encompass the decay products ($0<y<\mnus x/2$), and for
the BBN calculations we use $\eta_{10}=5$.

The results of the calculations have been imported into the modified
Kawano nucleosynthesis code~\cite{kawano}, and the abundances of all
light elements have been calculated.  At each time step $x$ we find
the corresponding photon temperature and total energy
density. Furthermore we integrate the kinetic equation governing the
$n/p$ evolution taking into account the distorted spectrum of $\nue$.

For a given mass, $\mnus$, we calculate the helium abundance, $Y$, for
various mixing angles (or lifetimes) 
and use the approximate formula $\Delta Y = 0.013 \, \Delta N$ 
to find the minimum allowed mixing angle (Fig.~\ref{fig:smax1}) or the
maximum allowed lifetime (Fig.~\ref{fig:tmax1})
for that mass.  We
do this for both $\Delta N =0.2$ and $1.0$.
On the same figures we plot guide-the-eye lines, 
that are fitted in
the form 
\be
(\sin^2 \theta)_{\rm min} &=& s_1\,M_s^\alpha +s_2 
\label{sin}
\ee
for the minimum allowed mixing 
 and
\be
\tau_{\rm max} = t_1\,M_s^\beta +t_2
\label{time}
\ee
for the maximum allowed life-time $\tau_{\rm max}$. 
Fitting parameters in Eqs.(\ref{sin},\ref{time}) are presented 
at the Table \ref{tab1}.

\begin{center}
\begin{table}[ht]
\begin{center}
\begin{tabular}{|c|c|c|c|c|c|c|c|} 
 \hline\hline
&&\multicolumn{3}{|c|}{~}&\multicolumn{3}{|c|}{~}\\
~{\bf mixing}~&~~${\bf \Delta N}$~~&
\multicolumn{3}{|c|}{$(\sin^2 \theta)_{\rm min}$ fitting parameters.}&
\multicolumn{3}{|c|}{$\tau_{\rm max}$ fitting parameters.}\\
\cline{3-5}\cline{6-8}
&&~~${\bf s_1}$~~&~~${\bf s_2}$~~&~~${\bf \alpha}$~~ &~~${\bf t_1}$~~&~~${\bf t_2}$~~&~~${\bf \beta}$~~\\
\hline
&&&&&&&\\
$\nu_s-\nu_{\mu,\tau}$&$0.2$&$1423$&$-4.99 \cdot 10^{-6}$&$-3.727$&$100.1$&$0.04788$&$-1.901$\\
&&&&&&&\\
\hline
&&&&&&&\\
$\nu_s-\nu_{\mu,\tau}$&$1$&$568.4$&$-5.17 \cdot 10^{-6}$&$-3.549$&$128.7$&$0.04179$&$-1.828$\\
&&&&&&&\\
\hline
&&&&&&&\\
$\nu_s-\nu_{e}$&$0.2$&$140.4$&$-9.9\cdot 10^{-6}$&$-3.222$&$1218$&$0.0513$&$-2.658$\\
&&&&&&&\\
\hline
&&&&&&&\\
$\nu_s-\nu_{e}$&$1$&$66.33$&$-1.05\cdot 10^{-5}$&$-3.070$&$1699$&$0.0544$&$-2.652$\\
&&&&&&&\\
\hline
\hline
\end{tabular}
\end{center}
\caption{Fitting parameters for minimum allowed mixing $(\sin^2 \theta)_{\rm min}$ fitted by Eq.(\ref{sin})
and maximum allowed life-time $\tau_{\rm max}$ fitted by Eq.(\ref{time}).}
\label{tab1}
\end{table}
\end{center}

For small masses we followed Ref.~\cite{dolgov00} and used the
expansion in a small parameter, $\Delta=T \cdot x -1$, to describe
the temperature evolution, whereas for bigger masses we used a more
correct, and somewhat more complicated, treatment of the photon
temperature from Eq.~(\ref{energ}).  The results are in perfect
agreement for $\mnus=60$~MeV, and for $\mnus=140$~MeV we find the
maximal allowed lifetime about $0.05$ sec, in fair agreement with the
approximate fitting formulae.

It is important to keep in mind, that the bounds obtained above are
rather conservative, since all the approximations used in the
derivation lead to slightly weakened bounds, as described in
Ref.~\cite{dolgov00}.

\section{Decay $\nu_2 \rar \nu_a +\pi^0$}

The results obtained in the previous section are valid when the mass
of the heavy neutrino is 
smaller than 140 MeV. For such low masses the dominant decay
modes of $\nu_2$ would be into electrons and light neutrinos. A 
possible decay mode including muons which is possible for $M_s > 105$ MeV
is suppressed if $M_s < 140$ MeV. However, for $M_s > 135$ MeV the decay
channel
\be
\nu_2 \rar \pi^0 + \nu_a
\label{decaypi}
\ee
becomes open and strongly dominating. The life-time of 
$\pi^0 \rar \nu \bar \nu $ was calculated in 
refs.~\cite{fishbach77,kalogeropoulous79}. We can translate their
results for the decay (\ref{decaypi}) and find for the life-time
\be
\tau = \left[{ G_F^2 M_s (M_s^2 -m_\pi^2) f_\pi^2 \sin^2\theta 
\over 16\pi} \right]^{-1}
= 5.8 \cdot 10^{-9} \,{\rm sec} \left[ \sin^2 \theta { M_s (M_s^2-m_\pi^2)
\over m_\pi^3} \right]^{-1}
\label{taupi}
\ee
where $m_\pi = 135$ MeV is the $\pi^0$-mass and $f_\pi= 131 $ MeV is
the coupling constant for the decay $\pi^+ \rar \mu+\nu_\mu$. 

Immediately after $M_s$ becomes bigger than $m_\pi$ the two-body 
decay becomes the main one and all other channels can be neglected.
We cannot directly apply our numerical program (that was made for a
33.9-MeV neutrino) to this case, so we will
instead make some simple order of magnitude estimates
for the impact of very heavy $\nu_2$ on BBN. One can roughly conclude
that for the life-time of $\nu_2$ smaller than 0.1 sec, and corresponding
cosmological temperature higher than 3 MeV, the decay products would
quickly thermalize and their impact on BBN would be small. For a life-time
larger than 0.1 sec, and $T<3$ MeV, we assume that thermalization of
neutrinos is negligible and approximately evaluate their impact on BBN.
If $\nu_s$ is mixed with $\num$ or $\nut$ then electronic neutrinos are
not produced in the decay (\ref{decaypi}) and only the contribution of the
decay products into the total energy density is essential. As we have
already mentioned, non-equilibrium $\nue$ produced by the decay would
directly change the frozen $n/p$-ratio. This case is more complicated
and demands a more refined treatment than what is presented in this
section.

The $\pi^0$ produced in the decay (\ref{decaypi}) immediately decays  
into two photons and they heat up the electromagnetic part of the plasma, 
while neutrinos by assumption are decoupled from it. We estimate the
fraction of energy delivered into the electromagnetic and neutrino
components of the cosmic plasma in the instant decay approximation.
Let $r_s=n_s/n_0 $ be the ratio of the number densities of the heavy  
neutrinos with respect to the equilibrium light ones, 
$n_0 = 0.09 T_\gamma^3$. The total energy of
photons and $e^+e^-$-pairs including the photons produced by the
decay is
\be
\rho_{em} = {11\over 2}{\pi^2 \over 30} T^4 + r_s n_0 {M_s \over 2} 
\left( 1+{m_\pi^2 \over M_s^2}\right)\, ,
\label{rhoem}
\ee
while the energy density of neutrinos is
\be
\rho_{\nu} = {21\over 4}{\pi^2 \over 30} T^4 + r_s n_0 {M_s \over 2} 
\left( 1- {m_\pi^2 \over M_s^2}\right) \,.
\label{rhonu}
\ee
The effective number of neutrino species at BBN can be defined as
\be
K_\nu^{(eff)} = {22\over 7} {\rho_\nu \over \rho_{em}} \, .
\label{keffnu}
\ee
Because of the stronger heating of the electromagnetic component of the
plasma by the decay products, the relative role of neutrinos diminishes
and $K_\nu^{(eff)}$ becomes considerably smaller than 3. If $\nu_s$ are
decoupled while relativistic their fractional number would be
$r_s=4$ (two spin states and antiparticles are included). Possible
entropy dilution could diminish it to slightly below 1. Assuming that the 
decoupling temperature is $T_d = 3$ MeV we find that 
$K_\nu^{(eff)} =0.6 $ for $M_s = 150$ MeV and 
$K_\nu^{(eff)} =1.3 $ for $M_s = 200$ MeV if the frozen number density
of $\nu_s$ is not diluted by the later entropy release and $r_s$ remains
equal to 4. If it was diluted down to 1, then the numbers would 
respectively change to $K_\nu^{(eff)} =1.15 $ for $M_s = 150$ MeV and 
$K_\nu^{(eff)} =1.7 $ for $M_s = 200$ MeV, instead of the standard
$K_\nu^{(eff)} =3 $. 
Thus a very heavy $\nu_s$ would result in under-production of $^4 He$.
There could, however, be some other effects acting in the opposite direction.

Since $\nue$ decouples from electrons/positrons at smaller temperature
than $\num$ and $\nut$, the former may have enough time to thermalize.
In this case the temperatures of $\nue$ and photons would be the same
(before $e^+e^-$-annihilation) and the results obtained above would be
directly applicable. However, 
if thermalization between $\nue$ and $e^\pm$ was not efficient, then 
the temperature of electronic neutrinos at BBN would be smaller than in
the standard model. The deficit of $\nue$ would produce an opposite
effect, namely enlarging the production of primordial $^4 He$, because it
results in an increase of the $n/p$-freezing temperature. This effect
significantly dominates the decrease of $K_\nu^{(eff)}$ discussed above.
Moreover even in the case of the decay 
$\nu_2 \rar \pi^0 + \nu_{\mu, \tau}$, when $\nue$ are not directly 
created through the decay,
the spectrum of the latter may be distorted at the high energy tail
by the interactions with
non-equilibrium $\nut$ and $\num$ produced by the decay. This would
result in a further increase of $^4He$-production. In the case of 
direct production of non-equilibrium $\nue$ through the decay
$\nu_2 \rar \pi^0 + \nu_e$ their impact on $n/p$ ratio would be even 
much stronger. 

To summarise, there are several different effects from the decay of $\nu_s$
into $\pi^0$ and $\nu$ on BBN. Depending upon the decay life-time
and the channel these effects may operate in opposite directions. 
If the life-time of $\nu_2$ is  larger  than 0.1 sec but smaller than
0.2 sec, so that $e^\pm$ and $\nue$ establish equilibrium the production
of $^4He$ is considerably diminished so that this life-time interval
would be mostly forbidden. For life-times larger than 0.2 sec the
dominant effect is a decrease of the energy density of $\nue$ and this
results in a strong increase of the mass fraction of $^4 He$. Thus large 
life-times should also be forbidden. Of course there is a 
small part of the parameter space where both effects cancel each other
and this interval of mass/mixing is allowed. It is, however, difficult to
establish its precise position with the approximate arguments
described above. It would be a matter of separate and rather complicated
non-equilibrium calculations.

Thus, in the case of $\nu_s \leftrightarrow \nu_{\mu, \tau}$ mixing for 
$M_s>140$ MeV we can exclude the life-times of 
$\nu_s$ roughly larger than 
0.1~sec, except for a small region near
0.2~sec where two opposite effects cancel and 
the BBN results remain undisturbed despite the presence of sterile neutrinos.
Translating these results into mixing angle according to
Eq.(\ref{taupi}), we conclude that mixing angles 
$\sin^2 \theta < 5.8 \cdot 10^{-8} m_\pi/M_s
/((M_s/m_\pi)^2-1)$
are excluded by BBN. Combining this result with Eq.(\ref{dmsin})
we obtain the exclusion region for  $M_s>140 MeV$:
\be
5.1 \cdot 10^{-8} ~\frac{\mbox{MeV}^2}{M^2} 
<\sin^2 \theta < 5.8 \cdot 10^{-8}
~\frac{m_\pi}{M_s}\frac{1}{(M_s/m_\pi)^2-1}~.
\label{sin_pi_excl}
\ee 
   
In the case of $\nu_s \leftrightarrow \nu_e$ mixing 
for $M_s>140 MeV$ the limits are possibly stronger,
but it is more difficult to obtain reliable estimates
because of a strong influence of non-equilibrium $\nu_e$ produced by the
decay on neutron-proton reactions, and
we will postpone this problem for future investigations.

\section{Supernova 1987A}

In Ref.~\cite{dolgov00} we have explained that massive
sterile neutrinos can be constrained by the duration of the SN 1987A
neutrino burst. If the sterile states live long enough and
interact weakly enough to escape from the SN core, the usual
energy-loss argument constrains the allowed interaction strength (in
our case the active-sterile mixing angle). If the sterile states are
so short lived or so strongly interacting that they decay or scatter
before leaving the SN core, they contribute to the transfer of energy
and thus still accelerate the cooling speed of the SN core.

The arguments given in Ref.~\cite{dolgov00} pertain directly to the
present case as long as the sterile neutrino mass is not too large for
its production to be suppressed in the relevant region of a SN
core. Taking a typical temperature to exceed 30~MeV, the average
thermal neutrino energy exceeds about 100~MeV so that threshold
effects should not become important until the mass significantly
exceeds this limit. Therefore, we believe that the SN limits pertain
at least to masses up 100~MeV. For such a mass and maximal mixing, the
lifetime is about $10~\mu{\rm s}$, for relativistic particles
corresponding to a distance of about 3~km, i.e.\ of order the SN core
radius, and much larger than the standard mean free path of active
neutrinos of a few meters. Therefore, for the entire range of
interesting masses and mixing angles the lifetime far exceeds what is
necessary for the sterile neutrinos to transfer energy within the SN
core in the spirit of the energy-transfer argument.

In summary, the arguments of Ref.~\cite{dolgov00} imply that for
sterile neutrino masses below about 100~MeV the approximate range
$3\times10^{-8}<\sin^2(2\theta)<0.1$ is excluded.

\section{Conclusion and Summary}

\begin{figure}
\begin{center}
\psfig{file=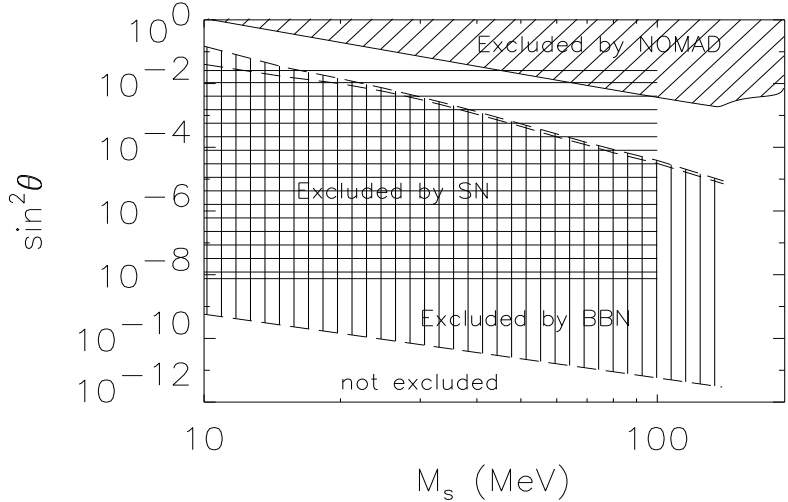,width=4.5in,height=3in}
\caption{Summary of our exclusion 
regions in the $\sin^2 \theta$-$M_s$-plane.  SN~1987A excludes all
mixing angles between two solid horizontal lines. BBN excludes the
area below the two upper dashed lines if the heavy neutrinos were abundant
in the early universe. These two upper dashed lines both correspond to 
the conservative limit of one extra light neutrino species permitted
by the primordial $^4$He-abundance. The higher of the two is for
$\nu_{\mu, \tau}$ mixing, and the slightly lower curve is for $\nu_e$ mixing.
In the region below the lowest dashed curve the
heavy neutrinos are not efficiently produced in the early universe and
their impact on BBN is weak. 
For comparison we have also presented here the region excluded by NOMAD
Collaboration~\cite{nomad} for the case of
$\nu_s \leftrightarrow \nu_\tau$ mixing.}
\label{fig:all}
\end{center}
\end{figure}

In this work we have found cosmological and astrophysical bounds on
possible mixings of a heavy sterile neutrino with mass $10~{\rm
MeV}~<~M_s~< 140~{\rm MeV}$ with any active flavour $\nue$,
$\num$ or $\nut$. Our results are summarised in
Fig.~\ref{fig:all}. The region between the two horizontal lines
running up to 100 MeV are excluded by the duration of the neutrino
burst from SN~1987A. A more detailed discussion would probably permit
one to expand this region both in the horizontal and vertical
directions.

The BBN limits are presented by the two upper dashed curves, both
corresponding to the conservative bound of one permitted extra
neutrino species. The curve for $\nu_{\mu, \tau}$ mixing is  slightly higher
than the curve for $\nu_e$ mixing.
The lower dashed curve describes our approximate estimate
of the efficiency of sterile neutrino production in the early universe
Eq.~(\ref{dmsin}).  Below this curve the heavy $\nu_2$ are very weakly
produced and their impact on BBN is not essential.  Let us note that
relation Eq.~(\ref{relat}) is fulfilled for the entire BBN-excluded
region of Fig.~\ref{fig:all}. In other words, sterile neutrinos are
relativistic at decoupling and the Boltzmann suppression factor is not
essential.  

We have also made an order of magnitude estimates of the 
BBN-influence of 
heavier neutrinos with masses $140~ \mbox{MeV} < M_S < 200 ~\mbox{MeV} $
for $\nu_s \leftrightarrow \nu_{\mu, \tau}$ mixing channel. 
We have found, that mixing angles in the range 
given by Eq.(\ref{sin_pi_excl})
are excluded by BBN. We do not present these order of magnitude
estimates in our Fig.~\ref{fig:all} since they are less precise than
the limits presented for $M_s < 140$~MeV.
We believe that the BBN bounds also could be improved
considerably with more detailed calculations.
For the $\nu_s \leftrightarrow  \nu_\tau$ mixing a part of the
$\sin^2 \theta$-$M_s$-plane is excluded by the NOMAD 
experiment~\cite{nomad}.
Their bound is also presented in Fig.~\ref{fig:all}. 

We have done the calculations for the Dirac neutrinos. For
the Majorana case the cosmological number density of the heavy
neutrinos would be twice smaller and corresponding bounds would be
changed in an evident way and rather weakly.

It is noteworthy that the impact of sterile neutrinos on BBN is very
sensitive to the deviation from thermal equilibrium of both sterile
neutrinos as well as their decay products, especially into the
$\nue$-channel. In the usual approximate calculations, kinetic
equilibrium is assumed. This assumption very much simplifies the
calculations because the integro-differential kinetic equations are
reduced to simple ordinary differential equations. However, we have
found that this approximation deviates substantially from the more
exact treatment presented here.  Even more accurate calculations are
feasible with the technique developed in our previous
works~\cite{dolgov98,dolgov99,dolgov97} where an accuracy at the
sub-percent level was achieved, but this is a difficult and
time-consuming problem.  We would like to postpone it until the
existence of heavy sterile neutrinos becomes more plausible.

\section*{Acknowledgement}
We are grateful to NOMAD collaboration for the permission to
present their results before publication.
We thank S.~Gninenko for stimulating our interest to this problem,
for numerous discussions, and for the indication of the importance
of $\nu_2 \rightarrow \pi^0 \nu_a$-decay.
A.~Dolgov is grateful to the Theory Division of CERN for the
hospitality when this work was completed. In Munich, this work was
partly supported by the Deut\-sche For\-schungs\-ge\-mein\-schaft
under grant No.\ SFB 375. The work of DS supported in part by INTAS
grant 1A-1065.

\end{document}